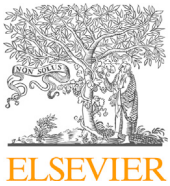
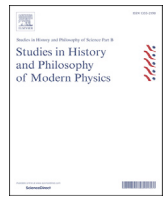

# Three problems about multi-scale modelling in cosmology

Michela Massimi[*]

School of Philosophy, Psychology and Language Sciences, The University of Edinburgh, Dugald Stewart Building 3 Charles Street, Edinburgh EH8 9AD, UK

## 1. Introduction

In the vast literature on scientific models and simulations, increasing attention has been paid to the sensitivity of modelling to scales. Models and simulations are tied to particular scales, with far-reaching consequences for the ontological inferences that can be drawn about the target system. For example, fluid dynamics typically offers scale-based models for the behavior of fluids. It is possible to model a fluid at the macro-scale (e.g. its viscosity) without having to model what happens at the micro-scale (and the Navier-Stokes equations typically take care of this modelling task at the macro-scale). On the other hand, if the goal is to model and understand the statistical behavior of molecules composing the fluid, then modelling at the macro-scale does not help with the task. Batterman (2013) has called this well-known feature of modelling practices the "tyranny of scales"; and has addressed the problem by looking at the role of the renormalization group in modelling what he calls the "in between scales" point of view. Batterman has explored methodological ways of modelling "in between" scales as a way of counteracting the tyranny of scales and the apparent choice it forces upon us between pure top-down or pure bottom-up modelling techniques.[1] Batterman's message is that "mesoscopic structures cannot be ignored and, in fact, provide the bridges that allow us to model across scales" because they are the scale at which top-down modelling strategies and bottom-up modelling strategies typically meet:

> bottom-up modelling of systems that exist across a large range of scales is not sufficient to yield observed properties of those systems at higher scale. Neither is complete top-down modelling. After all, we know that the parameters appearing in continuum models must depend upon details at lower scale levels. The interplay between the two strategies—a kind of mutual adjustment in which lower scale physics informs upper scale models and upper scale physics corrects lower scale models—is complex, fascinating and unavoidable (Batterman (2013), p. 283).

The overarching goal of the present paper is to contribute to this literature of modelling across scales by looking not at hydrodynamics, or condensed matter physics, but instead at contemporary cosmology. This paper explores the distinctive ways in which the "tyranny of scales" affects modelling dark matter at the meso-scale as I am going to call the scale of individual galaxies in contemporary cosmology. Three scale-related problems about modelling dark matter are presented and discussed. These problems are interesting because they lie at the very heart of cutting-edge research in cosmology; and, they touch upon pressing methodological aspects on which the debate in cosmology between the standard model (ΛCDM) vs. Modified Newtonian Dynamics (MOND) has been revolving. I am going to call them the "downscaling problem"; the "upscaling problem"; and the "in between" scales problem. Before presenting each of them, it is necessary to flesh out both the philosophical and scientific context behind the problem of multiscale modelling in contemporary cosmology.

What Batterman calls the "tyranny of scales" affects modelling in cosmology in interesting and novel ways, I am going to argue. For it affects the very nature of the debate between the ΛCDM model and MOND as well as some novel hybrid proposals that have emerged to go beyond the ΛCDM − MOND dichotomy. The ΛCDM − MOND debate has been ongoing since the 1980s when the work of Milgrom (1983) challenged the standard ΛCDM model, which postulates the existence of cold dark matter, and suggested instead the need for amending the laws of Newtonian dynamics at large scale. While the majority of the cosmology community has gathered consensus around the ΛCDM model —because of its remarkable success at explaining structure formation at large scale, among many other phenomena (as explained in Section 2)— as of today, no dark matter particles have been detected yet.

---

* Corresponding author.
  E-mail address: michela.massimi@ed.ac.uk.

[1] In Batterman's (2013, p. 256–257) own words, a reductionist view might assume for example that "whatever the fundamental theory is at the smallest, basic scale, it will be sufficient in principle to tell us about the behavior of the system at all scales. Continuum modelling on this view represents an idealization—as Feynman has said, 'a smoothed-out imitation of a really much more complicated microscopic world' (Feynman, Leighton and Sands 1964, p. 12) … Many believe that a reductionist/eliminativist picture is the correct one. I maintain that even if we can explain the safety and robustness of continuum modeling (how this can be done is the focus of this essay), the reductionist picture is mistaken. It is a mistaken picture of how science works. My focus here is on a philosophical investigation that is true to the actual modeling practices of scientists (…). The fact of the matter is that scientists do not model the macroscale behaviors of materials using pure bottom-up techniques. I suggest that much philosophical confusion about reduction, emergence, atomism, and antirealism follows from the absolute choice between bottom-up and top-down modeling that the tyranny of scales apparently forces upon us".

There is more. In the past two decades increasingly more accurate astrophysical data and measurements concerning galaxies have been made. These data have highlighted some phenomena—for example, the so-called Baryonic Tully-Fisher relation, (BTF hereafter) among others—that naturally fit MOND but are more problematic to explain within the ΛCDM model. Most within the cosmology community would not be swayed by BTF evidence at galactic scale, while MOND supporters have increasingly pointed at this evidence to keep the debate open. It is fair to say that many in the cosmology community feel that this debate will be settled if and when either dark matter is detected in the laboratory; or when MOND can be generalized from the status of a successful phenomenology at the galaxy scale onto a general theory at large scale.

From a philosophical point of view, however, this debate goes deeper than the dichotomy between finding the smoking gun of dark matter particles; or elaborating a suitable relativistic MOND for large-scale structure. What is philosophically at stake in this debate is yet another manifestation of the "tyranny of scales"; or better, how challenging multi-scale modelling proves when the scales in question are of the order of cosmological scales. My overarching philosophical goal is to lay out the nature, problems, and prospects of multi-scale modelling in contemporary cosmology by advancing and substantiating five main claims:

(i) ΛCDM and MOND, respectively, work best at a specific scale (large scale for ΛCDM, meso or galactic scale for MOND);
(ii) Each model faces challenges when modelling across more than one scale. The ΛCDM faces problems going down from large scale structure formation to the meso scale of individual galaxies—I call this the "downscaling problem". MOND faces problems going up from the meso scale of individual galaxies to the large scale of clusters and structure formation—I call this the "upscaling problem".
(iii) Philosophically, these problems are different in nature and different physical solutions to them have been given. The downscaling problem for ΛCDM is a problem about the *explanatory* power of ΛCDM for some recalcitrant phenomena at the meso scale of individual galaxies. The upscaling problem for MOND, by contrast, is a problem of *consistency*: how to consistently extend MOND at large scale where general relativity (GR) applies (and most of the large scale phenomena and even experimental techniques, such as gravitational lensing, rely on and presuppose the validity of GR).
(iv) Hybrid models have recently been explored (e.g. dark matter superfluidity discussed in Section 5). These models aim to bypass the ΛCDM − MOND dichotomy by delivering the best of both worlds at the large scale and the meso scale. They too, I contend, face challenges: I call it the "in between" scales problem. And it is not a problem about explanatory power, or consistency, but about the *predictive novelty* of the hybrid model and the extent to which its success is genuinely independent of the success of either ΛCDM or MOND.
(v) Ultimately, I suggest, a successful multi-scale cosmology ought to address these problems of scales that different models respectively face. Even if dark matter particles were detected tomorrow, the ΛCDM would still need to come up with an explanation of the BTF relation for galaxies. Even if MOND were able to retrieve clusters' behavior, the problem of consistency with large scale phenomena (CMB angular power spectrum, the matter spectrum, gravitational lensing) would still have to be addressed.

The paper is structured as follows. Section 2 introduces the standard ΛCDM cosmological model, and very briefly reviews the evidence for it at large scale and ongoing searches for dark matter particles. The emphasis in this Section is on the meso-scale of individual galaxies (rather than the micro scale of dark matter particle candidates, or the large scale of structure formation in the universe). This is the scale at which most of the contemporary debates and controversy takes place. Three problems about multiscale modelling are presented. Section 3 illustrates what I refer to as the "downscaling problem", faced by the current cosmological model, with a focus on some very recent developments on the front of computer simulations and their ability to tackle this problem. Section 4 discusses the "upscaling problem" faced by modified Newtonian Dynamics (MOND), as a rival of the current cosmological model. I review in particular two recent attempts at expanding upon MOND (the so-called EMOND and Verlinde's emergent gravity). Section 5 briefly illustrates a recent proposal that has been put forward with an eye to overcoming the stand-off between the downscaling and the upscaling problem. This hybrid proposal (by Berezhiani and Khoury) while exciting and promising, faces its own problem, namely what I call the "in between" scale problem, in the creative attempt at devising modelling solutions for dark matter across different scales. In the final Section 6, I draw some qualified and tentative conclusions on the philosophical and physical challenges still open for multiscale modelling in contemporary cosmology.

## 2. Modelling dark matter

According to the current cosmological model (the so-called "concordance model", or ΛCDM model), the universe consists of 70% dark energy, 25% dark matter and 5% ordinary matter. Clarifying the nature of dark matter and dark energy remains an open and pressing question for contemporary research both in particle physics and cosmology. Dark matter is the focus of the present paper (questions about dark energy would deserve an article in its own right and are therefore left on one side for the purpose of the present paper and its overarching goal).

The best evidence that there is dark matter in the universe traditionally comes from large-scale structure and galaxy clusters. But modelling dark matter at the macro-scale is not necessarily very informative about its micro-scale behavior, or its behavior at the meso-scale. Thus, modelling dark matter presents fascinating challenges when it comes to multi-scale modelling; challenges that have not been investigated so far in the otherwise vast literature on modelling and simulations (e.g., Morrison, 2009, 2015; Parker, 2009; Winsberg, 2010).

How to move from the micro-scale of, say, LHC physics (with scientists at CERN looking for possible candidates for dark matter among supersymmetric particles), to the large-scale structure of galaxies (studied by large cosmological surveys such as Planck)? Can bottom-up modelling techniques and top-down modelling techniques successfully meet at the meso-scale in the case of dark matter? To echo Batterman, what are the mesoscopic structures in this case that might provide the bridges to model dark matter across scales? This Section offers an overview of these challenges, whereas Sections 3, 4, and 5 zoom in the specific downscaling, upscaling, and in between scales problems, respectively.

The best and more compelling evidence as of today for dark matter comes from large-scale structure. It is at this scale that four main factors strongly point to the existence of dark matter. First, the need to reconcile the expansion rate of a geometrically flat Robertson-Walker universe with Big Bang Nucleosynthesis (BBN). Second, the angular power spectrum of the cosmic microwave background (CMB). Third, the matter power spectrum related to baryon acoustic oscillations (BAO) observed in the CMB. Fourth, the formation of galaxy clusters. Let us very briefly review each of these main evidential factors for cold dark matter.

The universe has long been known to be expanding with the Hubble parameter $H_0$ measuring the accelerated expansion (via Supernova Ia and other probes).[2] The matter density of the universe has been estimated to be around $\Omega_m \approx 1/4$ (i.e. 0.25) since the 1990s (see White, Navarro, Evrard, & Frenk, 1993 for example). A universe with a geometrically flat metric and a matter density less than 1 invited the introduction of a new parameter, namely dark energy (in the form of Einstein's cosmological constant $\Lambda$) that added to the matter density could add up to 1. Dark energy (whose nature remains to be understood and would deserve a paper in its own right) counterbalances gravity in the formation of structure and causes the universe to accelerate in its expansion. Out of the total matter density $\Omega_m \approx 0.25$, it turns out that only a small fraction is made up of baryons with the baryon density estimated around $\Omega_b \approx 0.05$. The baryon density is measured from the baryon-to-photon ratio. CMB provides an accurate indication of the photon energy density at the time of the last scattering, while Big Bang Nucleosynthesis (BBN) provides constraints on the abundance ratios of primordial elements (hydrogen, helium etc.) which formed after the Big Bang. The fact that the total matter density $\Omega_m$ far exceeds the baryon density $\Omega_b$ provides strong evidence for the existence of an additional kind of non-baryonic (maybe weakly interacting) matter which has not yet been observed: dark matter, precisely.

The second main factor in support of dark matter is the angular power spectrum of the cosmic microwave background (CMB), from maps such as the NASA/WMAP Science Team for cosmic microwave background and Planck (see Ade et al., Planck 2014; 2015), which show initial density fluctuations in the hot plasma at the time of last scattering (cf. Bennett et al., 2013; Hinshaw et al., 2013). The over-dense blue regions in these maps show the seeds that led to the growth of structure, and the gradual formation of galaxies and rich galaxy clusters (under the action of gravity) over time. Cosmologists infer the existence of a non-baryonic (weakly interacting) dark matter that must have not coupled with photons and be responsible for the structure formation at early stages as compatible with the observed angular power spectrum of the CMB. The growth of fluctuations from the CMB epoch to now has been of the order of $10^3$ while baryonic fluctuations in CMB are of a much smaller order than that, indeed too small to form galaxies. Structure formation can only be explained if one assumes dark matter whose fluctuations are $\sim 10$ times higher than in baryons.

Thirdly, evidence for dark matter (and evidence that cannot be similarly explained within MOND) comes from the matter power spectrum as observed in baryon acoustic oscillations (BAO). BAO are the remnants of original sound waves travelling at almost the speed of light shortly after the Big Bang and before the universe started cooling down and atoms formed (around 480 million light-years ago). This phenomenon resulted in the formation of what appears in the sky as an overdense region of galaxies forming a ring with radius of 480 million light-years from a given galaxy. By knowing the radius of the ring (which is a 'standard ruler' — i.e. 480 million light-years), cosmologists can measure the angle subtended from the Earth vantage point. BAO measurements are then used to probe how fast the universe has been accelerating at different epochs (and hence they are used as a probe for dark energy in this respect). But BAO are also important for dark matter because the estimated amplitude of their matter spectrum diverges wildly in a dark-matter model and in a no-dark matter model. In the former, such oscillations would soon fade away, i.e. the amplitude of the matter spectrum will not stay constant because baryon falling into the potential wells of dark matter. But in a no-dark matter model, one would expect the amplitude of such matter spectrum to remain fairly stable (which disagrees with BAO data collected by the Sloan Digital Sky Survey; see Dodelson, 2011 for a discussion).

Finally, using lensing techniques it is possible to study how large numbers of galaxies 'cluster together' and are much closer to each other than one would expect on the basis of gravity alone, and dark matter has traditionally been assumed to explain this phenomenon (cf. Bradley et al., 2008 for a recent study).[3] Needless to say, lensing itself assumes general relativity and since MOND does not, there is no MOND-like equivalent of this very same technique.

Turning from the large to the micro scale, on the other hand, particle physicists have been devising experiments for the direct detection of possible dark matter candidates. The favored candidates are hypothetical WIMPs (or weakly interacting massive particles), whose weak interaction with ordinary matter could lead to the recoils of atomic nuclei detectable using large liquid xenon chambers located underground. One such possible WIMP candidate is the so-called neutralino, the 'lightest supersymmetric particle' (LSP) whose searches at the Large Hadron Collider[4] (CERN), among other experiments, have given null results as of today. Similarly, direct detection of dark matter candidates at two of the largest experiments, such as LUX in South Dakota, and PandaX-II in China JinPing underground laboratory, has produced null results so far.[5] Alternative possible candidates for dark matter are hypothetical particles called axions,[6] gravitinos,[7] self-interacting dark matter (SIDM), and hypothetical superheavy and super-weakly-interactive particles called WIMPzilla,[8] among others. It goes beyond the goal and scope of this paper to investigate the plurality of existent hypotheses about dark matter candidates. For my goal is instead to explore how dark matter is typically modelled in between these two scales (i.e. the large cosmological scale of structure formation, and the micro-scale of particle physics).

---

[2] Ade et al. Planck (2015) performed an indirect and model-dependent measurement of the Hubble parameter based on $\Lambda$CDM and cosmic microwave background (CMB), which with more recent improvements (due to an increase in number of Supernova Ia calibrated by Cepheids, see Riess et al., 2016) has led to an estimated measurement value for the Hubble parameter of $73.24 \pm 1.74$ Mpc$^{-1}$ km/s. This value is in $3.4\sigma$ tension with the latest news from Aghanim et al. Planck (2016).

[3] Indeed, the term "dark matter" was originally introduced in the 1930s by the Swiss cosmologist Zwicky (1933) precisely to account for the phenomenon of the Coma galaxy cluster. But the very notion of dark matter did not take off in the cosmological community until much later, in the 1970s, when the idea of an invisible (or 'dark') component resurfaced to explain another puzzling phenomenon, namely the reason why spiral galaxies did not seemingly lose their distinctive shape by whirling (see Ostriker & Peebles, 1973). The hypothesis of a 'dark matter halo' was introduced to explain the phenomenon and the later measurements on spiral galaxies rotational velocities by Rubin and collaborators (Rubin, Ford, & Thonnard, 1980) corroborated Zwicky's original idea. For an insightful (albeit self-declared "personal" and "also necessarily biased") historical account of the history of dark matter, see Sanders (2010). For an alternative historical overview, more in line with the standard cosmological model, see Bertone and Hooper (2016).

[4] See ATLAS Collaboration (2015) and CMS Collaboration (2016) for some recent examples.

[5] See Tan et al., PandaX-II Collaboration (2016); and Akerib et al., LUX Collaboration (2017).

[6] See Okada and Shafi (2017); Di Vecchia, Rossi, Veneziano, and Yankielowicz (2017) just to mention two more recent examples.

[7] See Dudas, Gherghetta, Mambrini and Olive (2017).

[8] See Kolb and Long (2017).

Modelling dark matter at the level of individual galaxies (rather than at the scale of galaxy clusters, large-scale structure, or at the micro-scale of WIMPs, axions or else) is one of the open and pressing challenges for contemporary cosmology, and it has given rise to a debate about the very existence of dark matter. This is an example of how modelling at the meso-scale of individual galaxies becomes crucial in cosmology. But, one might ask, why call the scale of individual galaxies the "meso-scale"? In observational cosmology, the ΛCDM model is regarded as extremely good and accurate (and veridical, most cosmologists would add) in describing astrophysical observations at very large scale—i.e. the scale of horizon (~ 15,000 Mpc)—up to spacing of individual galaxies (~ 1 Mpc). Below the scale of ~ 1 Mpc, however, some astrophysical observations concerning individual galaxies have proved problematic for the ΛCDM model. Sometimes these problems for ΛCDM model are referred to as "small-scale issues with ΛCDM model",[9] where small-scales are defined as those occupied by galaxies that have a mass $M \leq 10^{10} M_\odot$ (where $M_\odot$ stands for solar mass) and a dark-matter halo with virial velocity $V_{vir} \leq 50 Km\,s^{-1}$. Considering that our Milky Way has a baryonic mass of the order of $10^{10} M_\odot$, the expression "small-scale" refers then usually to the domain of so-called dwarf galaxies (galaxies smaller than our own Milky Way), where some of the problems for the ΛCDM model are typically displayed.

However, I prefer to use the expression "meso-scale" (rather than "small-scale") in what follows (much as it is a small scale compared to galaxy-clusters and the scale of horizon!) for the following reasons. First, the specific problem for the ΛCDM model that I am going to focus on in this paper (i.e., the Baryonic Tully-Fisher relation) is not just displayed in dwarf galaxies. It is displayed also in our Milky Way and in galaxies bigger than the Milky Way. Hence, the reason for labelling individual galaxies (no matter what their solar mass or virial velocity or other parameters might be) as the "meso-scale". Second, in a more mundane sense, modelling dark matter in individual galaxies sits in between modelling dark matter at the microphysical scale (with WIMPs et al. candidates for cold dark matter) and modelling dark matter at the large-scale structure (with CMB and related structure formation).

The bone of contention at the meso-scale concerns the well-known observation that the rotational velocity of spiral galaxies instead of decreasing with distance from the center of the galaxy —as one would expect— is observed to remain flat. Given the centripetal gravitational acceleration responsible for the rotation of galaxies:

$$\frac{V^2}{r} = \frac{GM}{r^2}$$

if the observed velocity $V$ is constant one can infer that $M \propto r$. This is taken as evidence for the existence of dark matter halos surrounding galaxies, and inside which galaxies would have formed (the same massive halos, which incidentally, are necessary to guarantee dynamical stability to galactic disks). Dark matter halos have to be assumed to be dominant in the outskirt of the galaxies to explain the flat rotation curves. In a seminal paper, Julio Navarro, Carlos Frenk and Simon White (1996) laid the foundations of current cosmological research by studying the nature of the dark matter halos, inside which galaxies would presumably have formed after the Big Bang. This halo would have a distinctive "cusp" with dark matter density increasing towards the center, and it features prominently in all current cosmological simulations that deal—within the remit of the ΛCDM model—with the phenomenon of flat rotation curves for galaxies.

Yet critics of the ΛCDM model have for long time regarded this very same phenomenon of galaxies' rotation curves as evidence that the standard cosmological model might not be correct after all. The research programme known as MOND (or Modified Newtonian Dynamics, see Milgrom, 1983) explains galaxies' flat rotation curves by modifying Newton's acceleration law at the galaxy scales.[10] MOND supporters have long argued that it is possible to track how the observed rotational velocity of galaxies ($V$) exceeds the velocity that would be expected on the basis of Newtonian mechanics ($V_{Newton}$) as a function of the gravitational acceleration due to the baryonic content of the galaxies ($a_{Newton}$). Measuring the rotational velocities of a sufficiently large sample of galaxies, and plotting $V/V_{Newton}$ over $a_{Newton}$, an acceleration scale appears in the data (around $a_0 = 1.2 \times 10^{-10}\,ms^{-2}$) so that when $a_{Newton} \gg a_0$ the observed velocity coincides with the expected velocity on the basis of Newton's theory. But when $a_{Newton} \ll a_0$ the discrepancy between the observed and the expected velocity ($V/V_{Newton}$) increases significantly. This phenomenon (McGaugh, 1998; Sanders, 1990) takes the name MDAR or 'mass discrepancy–acceleration relation' and has recently attracted new attention among historians and philosophers of physics (for an excellent discussion, see Merritt, 2017 in the pages of this journal). A more precise way of expressing MDAR is that the ratio of total mass (including dark matter)–to–baryonic mass $M_{tot}/M_b$ at any given radius $r$ for each galaxy (which is behind the presumed discrepancy $V/V_{Newton}$) anti-correlates with the acceleration due to baryons $(a_{Newton})^{1/4}$ at low accelerations. This means that at low accelerations, where dark matter dominates and baryons are sub-dominant according to ΛCDM, the sub-dominant baryon content seems to surprisingly 'govern' how dark matter should distribute to reproduce the observed galaxies' rotational velocities. By contrast, MOND does not face this problem because it is designed to precisely account for the presence of such an acceleration scale $a_0$ in the data about galaxies' flat rotational velocities (by modifying Newton's acceleration laws below the relevant acceleration scale).

Within ΛCDM, MDAR can be explained by modelling dark matter halos that surround the baryonic–dominated core of galaxies (see Navarro et al. 2017). But modelling the standard Navarro–Frenk–White[11] (NFW) dark matter halo to fit MDAR may give too big halos. Too big dark matter halos face the so-called "too big to fail" problem and are not consistent with ΛCDM predictions (a problem first pointed out by Boylan-Kolchin, Bullock, & Kaplinghat, 2011). More recently, Stacy McGaugh (2015a,b) has returned to this problem by stressing the limitations of using ΛCDM simulations (with compressed NFW dark matter halos) to retrieve the data

---

[9] See for example Boylan-Kolchin et al. (2011) for an extensive and detailed review of what they call 'small-scale issues' with the ΛCDM model, including the missing satellites, the cusp-core problem, the too-big-to-fail problem, among others.

[10] For an excellent review of the MOND paradigm with a detailed explanation of both general predictions afforded by MOND, supporting data from galaxy phenomenology as well as clusters, and the Bekenstein-Milgrom theory, please see Sanders and McGaugh (2002).

[11] The NFW profiles for cold dark matter is usually characterized by two parameters: what is called the virial mass ($M_{200}$) – which is the mass of the halo spheroid where the enclosed mean density is 200 times the critical density of the universe; and 'concentration' $c = r_{200}/r_s$ (where $r_{200}$ is the virial radius and $r_s$ is the radius of the halo).

about the measured rotation curves of the Milky Way. In McGaugh's own words (2015a, p. 16):

> How the MDAR comes to be remains a mystery. In the context of ΛCDM, we are obliged to imagine that this very uniform scaling relation somehow emerges from the chaotic process of feedback during galaxy formation. Alternatively, it could be that the appearance of a universal effective force law in galaxy data (the MDAR) is an indication of an actual modification of the force law (MOND: Milgrom, 1983).

To better assess this ongoing debate in cosmology and more recent data, however, there is one further piece of evidence at the meso-scale of individual galaxies that has become the real bone of contention between supporters of the ΛCDM model (i.e. the vast majority of the cosmology community) and ΛCDM critics: i.e., the Baryonic Tully-Fisher relation (BTF). While MDAR measures the mass discrepancy $M_{tot}/M_b$ within each galaxy, the Baryonic Tully-Fisher relation (Tully & Fisher, 1977) is a global empirical relation that tightly correlates the flat velocity of disk galaxies to the power of 4 with the galaxy's baryonic mass $M_b$: $M_b \propto V^4$.[12] BTF has traditionally been difficult to explain within ΛCDM. The correlation between baryonic mass and $V^4$ has been observed to hold very tightly (with a very small measured scatter) across a wide range of galaxies: from large galaxies—where the higher baryonic content might seem naturally related to the higher galaxy's flat velocity—to small dwarf galaxies, where the baryonic content is low and dark matter seems to dominate. In dwarf galaxies, the BTF relation deviates from the simple power law of $M_b \propto V^4$ but this is usually rectified by including cold gas contributions in addition to the stellar mass, since the former are prevalent in dwarf galaxies. What is then surprising about the BTF relation is that it is observationally found to apply with a very small scatter across a wide range of galaxies masses (see Fig. 1 from McGaugh, 2015a, for how the Milky Way compares to other galaxies in the BTF relation).

More recently, Lelli, McGaugh, and Schombert (2016) have sampled 118 disc galaxies (spiral and not) to study BTF with high quality data; and they have found all these galaxies to follow this simple power-law relation between baryonic mass and flat velocity to the power of 4. How to explain the small scatter of the BTF relation? And why is it that in dwarf galaxies, which are dark-matter-dominated, the BTF relation is still proportional to the baryonic matter, which is sub-dominant? An industry has flourished around devising simulations that can retrieve the BTF relation within the ΛCDM model. This is a cutting-edge area of cosmological research at the meso-scale, to which I turn my attention in the next Section 3.

On the other hand, in MOND the BTF relation is a natural consequence of modifying Newton's law at the cosmological scale as $F = ma\mu (a/a_0)$, with $\mu (a/a_0) = 1$ for large accelerations ($a_{Newton} \gg a_0$) where Newtonian mechanics applies; and $\mu (a/a_0) = a/a_0$ for small accelerations ($a_{Newton} \ll a_0$) where Newton's law must be modified. Galaxies' centripetal acceleration ($a = V^2/r$) can be equated with the MOND acceleration ($g_N = F/m = GM/r^2$) at the deep MOND regime ($a_{Newton} \ll a_0$ where $\mu (a/a_0) = a/a_0$ and $\mu\left(\frac{a}{a_0}\right)a = g_N$). Multiplying by $a_0$ both sides, one obtains

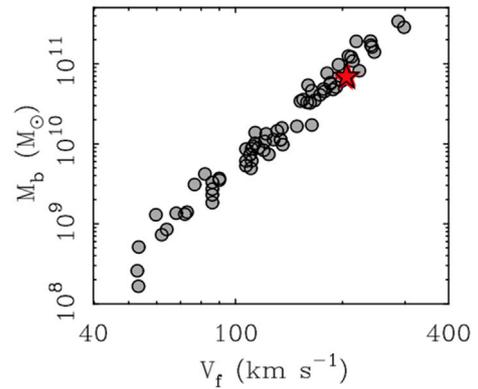

**Fig. 1.** The Baryonic Tully-Fisher relation that correlates galaxies's flat rotation velocity ($V_f$) with their baryonic mass ($M_b$) in solar masses ($M_\odot$). The Milky Way is the red star in this plot of spiral galaxies where BTF is observed to have a very small scatter. From S. McGaugh (2015a), p. 10. © AAS. Reproduced with permission.

$$a^2 = \left[\frac{V^2}{r}\right]^2 = a_0 \frac{GM}{r^2}$$

from which the BTF proportionality between baryonic mass and $V^4$ (i.e. $M_b \propto V^4$) naturally follows.

To sum up and conclude this Section, there is substantial disagreement at the meso-scale in contemporary cosmology:

1. The ΛCDM model (which is, and remains the official received view of the cosmological community at present) is able to retrieve these empirical facts about MDAR and the BTF relation by appealing to very complex 'feedback processes', which include stellar winds, supernova explosions, cooling and heating rates at play in early times galaxy formation simulations.
2. MOND (which is and remains a minority view in the cosmological community and is perceived as a successful phenomenology, but not more than that) offers a natural explanation for these empirical facts about MDAR and the BTF relation by rejecting dark matter and modifying acceleration laws below the threshold of a cosmological scale $a_0$. The underlying rationale is that if baryons dictate the behavior of galaxies' flat rotational velocities even in presumably dark-matter-dominated dwarf galaxies, this must be due to some anomalous properties of gravity (and not to some mysterious object called dark matter).
3. Recently, a number of theoretical proposals have been put forward that try to retrieve some of MOND empirical successes at the meso-scale (i.e. MDAR and BTF relation) and explain them in terms of dark matter behaving like a superfluid at low temperature (see Berezhiani & Khoury, 2015; Khoury, 2015). This third way promises to achieve the best of both worlds by explaining why MOND fares so well as a phenomenology at the meso-scale, but not at the large-scale structure.

Hence, three problems affect multi-scale modelling in contemporary cosmology:

I. *The downscaling problem* affects ΛCDM and its ability to *explain* (as opposed to just retrieve) the observed Baryonic Tully-Fisher relation at the level of individual galaxies.
II. *The upscaling problem* affects MOND and its ability to *consistently* retrieve large-scale structures, such as galaxy clusters, and structure formation at a scale where general relativity applies.

---

[12] This is BTF when the galaxy's velocity is flat (i.e. $V_{flat}$). Sometimes in the literature, one finds BTF expressed as a relation between baryonic mass and velocity to the power of 3.5, because a different velocity indicator is used (i.e. $V_{max}$). More in general BTF slopes range between the value of 3.5–4.1 depending on how stellar mass is estimated from observed light and dynamic range in the data. I thank Stacy McGaugh for helpful clarifications on this point.

III. The 'in between' scales problem affects hybrid models that currently try to achieve the best of both worlds (for example by treating dark matter as a superfluid) and this is a problem about the *predictive novelty* of the hybrid proposals.

In what follows, the problems and prospects of I.—III. are assessed by examining cutting edge research. Section 3 is dedicated to the downscaling problem; Section 4 to the upscaling problem; and Section 5 gives a short overview of the 'in between' scales problem. In the final Section 6, I return to the tyranny of scales and draw some preliminary conclusions about multiscale modelling in cosmology.

## 3. The downscaling problem for ΛCDM

Within ΛCDM, the anomaly of the BTF relation is typically retrieved by appealing to 'baryonic feedback', i.e. all the very complex baryonic physics involved in star and galaxy formation, ranging from active galactic nuclei (AGN, i.e. galaxies whose center is occupied by a black hole), to Supernovae explosions that can reduce the central dark matter density of the galaxy; from cooling rates to reionization processes, among others. Computer simulations within ΛCDM typically embed this complex baryonic feedback to successfully simulate galaxy formation at early epochs in a way that is consistent with/able to retrieve the observed BTF relation. Significant progress has been made in very recent years on this front to the point that it is fair to say that the majority of cosmologists nowadays do not regard the BTF relation any longer as an anomaly for the ΛCDM model. The results obtained by very recent computer simulations on this score are regarded by many in the community as the decisive sign that ΛCDM is on the right track (and it has been all along). From a philosophical perspective, this over-reliance on computer simulations to answer these pressing questions raises interesting questions, questions that this Section explores in some detail.

There has been a recent debate in philosophy of science about the increasing role of computer simulations and how they are changing the face of experimentation (Giere, 2009; Massimi & Bhimji, 2015; Morrison, 2009, 2015; Parker, 2009). While most of the debate has concentrated on the so-called materiality of ordinary experiments vs. non-materiality of computer simulations, there is a further distinctive element of computer simulations that so far has not attracted much philosophical attention; yet it plays a key role in cosmological computer simulations. I am going to call it the *context-sensitivity of the phenomenology* of computer simulations. Phenomenological models are well familiar in philosophy of science and ubiquitous in science (see Bokulich, 2011; Sterelny, 2006; Teller, 2004). Phenomenological models are regarded as ad hoc in fitting the model to the relevant empirical data and not necessarily explanatory in the sense of providing an explanation for the phenomenon in question.[13] Cosmological simulations within ΛCDM for the MDAR and BTF relations share precisely this phenomenological feature with more traditional phenomenological models used in ordinary experiments. There are two main kinds of cosmological simulations (for a discussion see Neistein, Khochfar, Dalla Vecchia, & Schaye, 2012): hydrodynamical simulations (HYDs) and semi-analytic simulations (SAMs). In both cases, there are free parameters, and the task of the simulation is to provide the best ad hoc fit of the free parameters to the relevant empirical data. This is done ad hoc by trying to reproduce key observables without having necessarily an algorithm or recipe that starting from first principles is able to deliver an explanation for the key observables. In other words, cosmological simulations calibrate a set of relevant free parameters to fit observables, as opposed to predicting or explaining the key observables themselves (this is what I called the *phenomenology* of computer simulation). But, even more importantly, this phenomenology proves deeply *context-sensitive*. And it is this context-sensitivity that underpins *the downscaling problem*, affecting the ability of ΛCDM to *explain* (as opposed to just retrieve) the observed Baryonic Tully-Fisher relation at the level of individual galaxies. Let us see why and how with a few detailed examples.

The main difference between hydrodynamical simulations and semi-analytic simulations is the following. HYD simulations model the evolution of galaxies over time by taking into account the complex hydrodynamical processes involved in the bottom-up process of growth of galaxies from dark matter seeds at early epochs. This process is affected by very complex baryonic physics (including cooling rates, gas accretion, Supernova feedback—i.e. heated gas that is ejected in explosions, among others), which are ultimately responsible for shaping the morphology of galaxies over time. HYD simulations have a finite resolution and to track this complex baryonic physics, they rely on 'sub-grid' physics. SAMs, on the other hand, are simpler to use because they describe the interaction of relevant parameters (such as the mass of stars, the mass of cold gas, the mass of hot gas, and black holes if any is present in AGN). SAMs offer laws for relevant processes involving these key parameters (e.g., laws for star formation, and cooling efficiency—i.e. how long it takes for the gas to reach the center of the dark matter halos and cool down; and so forth). HYD simulations model the behavior over time of a large number of particles (say, $10^6$–$10^9$ particles), behavior that is affected by a variety of context-sensitive factors (cooling rate, re-ionisation, supernova explosions, AGN, etc); any random deviation for any of these factors is going to affect the final galaxy morphology. SAM simulations, on the other hand, use only average efficiencies for some of these same processes per dark matter sub-halo and take the integrals of these efficiencies over time. In either case, by proceeding in this broadly phenomenological way (either via HYDs supplemented by sub-grid physics for feedback; or, via SAMs by analytically solving the relevant equations), cosmological simulations have come a very long way towards reproducing galaxy morphology and evolution at the meso-scale in a way that it was unimaginable until a decade ago. Cosmological simulations have been welcomed in the community as evidence that the ΛCDM model is indeed correct and capable of successfully addressing what I called above the downscaling problem. In the rest of this Section, I present some of the recent HYDs and SAMs that have successfully addressed the downscaling problem. The goal of this Section is not just to report on the state of the art in the field, but also to offer a philosophical angle that might prove helpful in assessing some of the open methodological issues in this debate.

One family of such recent cosmological hydrodynamical simulations at low resolution are the EAGLE and Apostle. In a recent paper, Ludlow et al. (2017) have used these simulations to show how—regardless of the feedback mechanism (weak or strong, AGN or not)—the total acceleration vis-à-vis baryonic acceleration matches the observed data even for models whose sub-grid physics was not tuned to match observational constraints (see Fig. 2).

However, despite this success, these same cosmological hydrodynamical simulations have proved problematic to account for the great diversity of shapes of rotation curves, especially for dwarf galaxies. Dwarf galaxies, with low-mass content and prevalently made of dark-matter, exhibit a great diversity of rotation curve shapes (even at fixed maximum rotation speed) that do not necessarily agree with the EAGLE simulations (as discussed in Oman et al., 2015).[14] In particular, some rotation curve shapes of these galaxies suggest a 'inner mass deficit': the inferred mass in the inner region of these dwarf galaxies is much lower than one would expect on the basis of ΛCDM simulations, even taking into account baryon-induced effects. This could be either evidence for a more complex model of dark matter. Or, as critics stress, evidence that cosmological hydrodynamical simulations simply fail to reproduce accurately the baryon effects in the inner regions of dwarf galaxies.

---

[13] In Alisa Bokulich's (2011) words: "A *phenomenological model* is only of instrumental value to the scientist. Often—though not exclusively—they are constructed via an ad hoc fitting of the model to the empirical data. Phenomenological models are useful for making predictions, but they do not purport to give us any genuine insight into the way the world is. An *explanatory model* on the other hand does aim to give us insight into the way the world is". On this ground, Bokulich argues for example that Bohr's model of the atom was not phenomenological, because it had a genuine explanatory import.

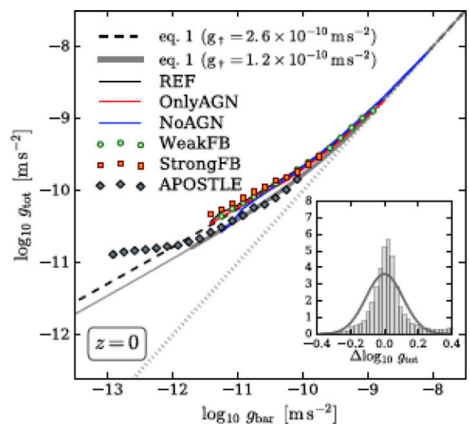

**Fig. 2.** Acceleration profiles for dark matter haloes given by APOSTLE (grey diamonds) as a function of baryonic acceleration and fit to different scenarios (with weak feedback in green circles; strong feedback in red squares; no AGN in blue line). Reprinted Fig. 3 with permission from Ludlow, Benítez-Llambay, Schaller, Theuns, Frenk, Bower, (2017). Mass-discrepancy acceleration relation: a natural outcome of galaxy formation in cold dark matter halos. *Physical Review Letters* 118: 161103., p. 4. Copyright (2017) by the American Physical Society.

On the SAMs side, Di Cintio and Lelli (2016), using semi-empirical models have been able to reproduce the BTF relation but with a scatter higher than the observed one (unless parameters are tuned accordingly). But the real turning-point in this literature has been a very recent article by Cattaneo et al. (2017) that has attracted significant attention in the community, with headlines along the lines of "dark matter theory triumphs" because of the unprecedented ability to retrieve both BTF relation and the Faber-Jackson relation (which is the equivalent of the BTF relation but not for spiral galaxies but for elliptical galaxies, in linking luminosity with stellar velocity dispersion). Let us take a closer look at how Cattaneo et al.'s new SAM—called GalICS 2.0—has been able to achieve such a remarkable result.

Cattaneo and collaborators have followed up on their previous work with SAMs, where it is standard practice to describe the formation and evolution of galaxies as a two-step process: (1) dark matter halos form from the primordial density fluctuations; (2) baryonic physics inside halos over time form stars and galaxies. Step (1) is modelled via merger trees for halos. Step (2) is broken down into a series of simplified processes whose differential equations can be solved analytically. What is interesting and peculiar about the semi-analytic GalICS 2.0 is that it treats the formation and evolution of galaxy in a modular way: different modules in the simulation take care of different aspects in both steps (1) and (2) at different scales. Thus, for example the TREE module takes care of the dark matter halo hierarchical formation (within the ΛCDM) using the TreeMaker algorithm to create a network of progenitor, descendent, host, sub-halos (where a halo is defined as descendent if it inherits more than half of its progenitor's particles). Halos are classified on the basis of some crucial properties such as virial mass, radius, angular momentum, but also concentration $c$, and position in the tree (progenitor/descendent), among others.

The module called HALO takes care of the baryonic physics concerning gas accretion, cooling rates etc. In particular, it controls all the cold gas and hot gas inflows and outflows in the halos. This is a crucial aspect of the GalICS 2.0 simulation and it is here that some key assumptions enter. For example, it is assumed that the cold gas accretion is the main mode of galaxy formation; and, moreover, that hot gas never cools down. These assumptions play a central role in making sure that GalICS2.0 produces results which are in agreement with observations. The authors acknowledge that these assumptions might be extreme and are justified on the basis that introducing a cooling term in the equations would only make sense if there were active galactic nuclei (AGN) to mitigate the cooling effects and "the physics of these models are uncertain" (Cattaneo et al., 2017, p. 1405). Thus, AGN feedback is intentionally not included in this SAM. Hence, "it is important to realise that, in this article, we are not arguing for the absence of cooling on physical ground. This is the simplest possible assumption within the cold-flow paradigm and we want to explore how far it can take us" (Cattaneo et al., 2017).

Then, there is the module called GALAXY, which computes all the morphological and kinematic properties of the galaxies and is articulated into two submodules: GALAXY EVOLVE describes the evolution of an individual galaxy over time; GALAXY MERGE studies morphological transformation caused by mergers. Crucially, the whole simulation relies on the relative independence between the HALO and the GALAXY modules.[15]

In the GALAXY module again assumptions enter in calculating disc radii and rotation speeds, because this SAM like others compute galaxy disc sizes by assuming that baryons and dark matter have the same angular momentum distribution and that angular momentum is also conserved, while knowing from cosmological hydrodynamical simulations that this is in fact not the case and angular momentum is not conserved. However, even if mistaken at the level of individual galaxies, the assumption is retained because of its statistical validity and agreement with observations (Cattaneo et al., 2017, p. 1405).

---

[14] It is worth stressing in the words of Oman et al. (2015) that "the rotation curves of *many* galaxies, dwarf included, are actually consistent with ΛCDM predictions. This is important to emphasize since it is often thought that ΛCDM rotation curves are in conflict with *all* or *a majority* of galaxies, especially dwarfs … Actually the main difference between simulated and observed rotation curves is the great *diversity* of the latter (especially for dwarfs) which is unexpected according to our results" Oman et al., 2015, p. 3655. This unexpected diversity of rotation curve shapes does not constitute a surprise for MOND supporters, who explain it as a consequence of MDAR and appeal to the shape of galaxy rotation curves in low surface brightness galaxies as the fundamental advantage of MOND over the standard model. I thank an anonymous reviewer for this journal for stressing this point.

[15] Cattaneo et al. (2017, 1402) use the following analogy to explain the independence between modules: "The relation between HALO and GALAXY can be compared to that between a mill and a baker. There is an exchange of matter both ways (inflows and outflows, flour and money) but the baker does not need to know whether the flour has been ground with a water or a wind mill. Neither does the millman about the baker's recipes. This philosophy explains some practical choices such as that of the time substeps in Section 3.4. A galaxy contains different components, such as a disk, a bugle or a bar (…), but star formation and feedback within a component are followed in the COMPONENT module. The lowest scale corresponds to the STAR (stellar evolution) and GAS (interstellar medium) modules."

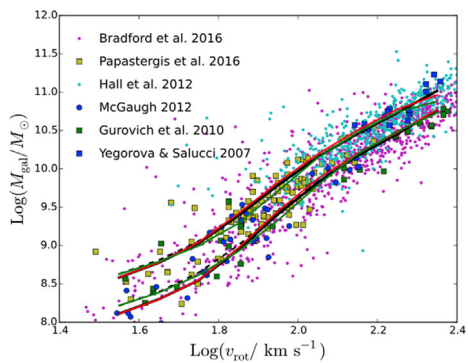

**Fig. 3.** BTF relation reproduced by GalICS2.0, where data points in different colors correspond to observed galaxies with data from McGaugh, 2012, Gurovich, Freeman, Jerjen, Staveley-Smith, and Puerari (2010) and so on. The color-coded curves correspond to the default model for GalICS2.0 in black solid, and three variations of it (labelled 'model 1, 2, and 3') in red, black dashed and green. © Cattaneo et al. (2017), p. 1420.

The module called COMPONENT describes star formation and feedback by tracking the exchange between gas in the interstellar medium and stars (whose formation is governed by another module called STAR). And stellar formation is hugely influenced by feedback at different scales. Supernovae explosions profoundly affect the dynamics of star formation and are extremely difficult to model because they are sensitive to contextual factors (i.e. if the supernova explodes inside a dense molecular cloud, most of the energy so released would be lost; photoelectrons in turn can inhibit star formation because they prevent overcondensation in dense molecular clouds). These feedback mechanisms normally play an important role in cosmological hydrodynamical simulations. But in GalICS2.0 "these complex physics is beyond the scope of our feedback model whose purpose is to compute the mass-loading factor (…), i.e. the rate at which cold gas is removed from galaxies. Any feedback mechanism that regulates star formation without removing gas from galaxies is already incorporated phenomenologically into our star formation efficiency" (Cattaneo et al., 2017, 1409).

In total, GalICS2.0 has then 16 free parameters, divided into various classes: *cosmological parameters* include matter density ($\Omega_m$), baryon density ($\Omega_b$), cosmological constant ($\Omega_\Lambda$) and the Hubble constant $H_0$; *dimensional parameters* including star formation threshold, SNe energy among others; *efficiency factors* such as disc instabilities, black hole accretion, and maximum SNe feedback efficiency; and *scaling exponents* such as $z$ scaling and $v_{vir}$ scaling. Large uncertainties affect some of these parameters such as the maximum SNe feedback efficiency, which are calibrated on galaxies in the local universe. Using this complex modular structure, GalICS2.0 has proved capable of retrieving the BTF relation (see Fig. 3 below), something that previous SAMs have not been able to.

Behind this unprecedented result, there is Cattaneo's definition of the rotational velocity of galaxies ($v_{rot}$) as being the same as the circular velocity ($v_c$) at the optical radius $r_{opt} = 3.2\,r_d$ "which contains 83% of the mass of an exponential disk" (Cattaneo et al., 2017, p. 1418). This is because "measuring $v_{rot}$ at the outer edge of the disk makes our results less sensitive to the real form of the dark matter density profile which is likely to differ at the center from the NFW model assumed in GalICS2.0" (Cattaneo et al., 2017). This assumption, however, seems to contrast with the way McGaugh (2012) data sets are gathered (the blue data point in Fig. 3), namely using "the rotation speed in the outermost regions, where the rotation curve is approximately flat" (as Cattaneo et al., 2017, 1421 acknowledges), even if the GalICS2.0 curves are broadly consistent with both McGaugh (2012) data sets and several others too. Moreover, the assumption that $v_{rot}$ is equal to the circular velocity ($v_c$) relies on the idealization that galaxy discs are "infinitely thin and cold so that the rotation speed $v_{rot}$ is equal to the circular velocity $v_c$ required to support a particle on a circular orbit against gravity" (Cattaneo et al., 2017, 1421).

Let us take stock. Where does this discussion about the use of computer simulations for modelling dark matter at the galaxy scale leave us? Despite there being an industry around dark matter simulations, few physicists in the field would claim that the success or failure of the ΛCDM model hangs on them. Instead, most physicists would probably claim that the problems pointed out above affect dark matter simulations no more than other kinds of computer simulations. And that no one should expect computer simulations to settle the debate between ΛCDM and MOND given how poor they might be at doing the job.[16] This common pragmatic attitude of the cosmology community about computer simulations is symptomatic of a more profound divide, which McGaugh (2015b, 253) has nicely captured: "If we are convinced that ΛCDM is correct, then modelling galaxies is a pesky distraction rather than a fundamental problem to address head on. Similarly, if we are convinced that MOND is correct, then ΛCDM is simply the best conventional proxy for the true cosmology of the underlying relativistic theory of MOND". Pesky distraction or not, the industry that has flourished around computer simulations in cosmology is also indicative that the problem with modelling galaxy phenomenology within ΛCDM cannot be brushed under the carpet.[17] As a philosopher, it is certainly not my goal or intention to try to settle the debate one way or another. Cosmology will settle the debate (as is to be expected). However, reflecting on the methodological assumptions and epistemic limits of scientific practice is the job of philosophers. And going back to the problem of multi-scale modelling in cosmology, here we touch a key point of this debate. In spite of its extraordinary success at explaining large-scale structure (i.e. structure formation, the matter power spectrum, galaxy clusters, and so on), ΛCDM is not equally well-equipped to *explain* phenomena such as BTF and MDAR at the scale of individual galaxies (what I have called the meso-scale). This scale has been traditionally regarded as favoring alternative models, such as MOND, which naturally *explains* BTF and MDAR because they are natural consequences of MOND formalism. Yet the task is not impossible for ΛCDM. As highlighted in this Section, the task has in fact been successfully addressed by recent work on HYD and SAM simulations for galaxy formation within the ΛCDM model.

However, as this Section has also shown, there is a cost for the ΛCDM model to pay in retrieving BTF and MDAR at the meso-scale. The cost is the appeal to very complex baryonic feedback in the process of modelling stars and galaxy formation. Although the baryonic physics behind feedback is well understood (i.e. photoionization, photoelectric heating, SNe explosions and so on), it proves nonetheless complex to model accurately baryonic physics in either HYD or SAM simulations. Simulation outcomes are very sensitive to small variations in how feedback is factored in, just to mention three examples:

---

[16] I thank an anonymous reviewer for pressing me to address this issue.

[17] For example, Sabine Hossenfelder's recent blog post on the topic is illuminating coming from someone who does not work in the field of computer simulations for dark matter: "the results of computer simulations are problem-ridden, and have been since the very first ones. The clumping matter, it turns out, creates too many small 'dwarfs' galaxies …. The simulations also leave some observations unexplained such as … the Tully-Fisher relation …. It is not something I used to worry about. … But after looking at a dozen or so papers, the problem Stacy [McGaugh] is going on about became apparent. … These papers typically start with a brief survey of other, previous, simulations, none of which got the structures right, all of which have been adapted over and over and over again to produce results that fit better the observations. It screams "epicycles" directly into your face" (Hossenfelder "Shut up and simulate (in which I try to understand how dark matter forms galaxies and end up very confused" full blog post in www.backreaction.blogspot.co.uk, 22 February 2018)

- whether or not AGN are included in the simulation;
- whether hot gas cooling might be mitigated by the presence of an AGN;
- whether SNe explosions take place in the center of a dense molecular cloud or not.

Even the best available simulations as of today are not capable of factoring in accurately all these competing (contributing and counteracting) causal factors, whose stochastic nature is very hard to model adequately, and which are nevertheless crucial ingredients in *explaining* individual galaxies' morphologies and evolution within ΛCDM.

The result is that even the best simulations available as of today either abstract and leave intentionally out some of these factors (as AGN in GalICS2.0); or idealize them (as with no-hot-gas-cooling in GalICS2.0); or mitigate large uncertainties by calibrating free parameters to a finite and limited sample of observations from the local Universe (as with the maximum SNe feedback efficiency in GalICS2.0, which is calibrated on the value of locally observable galaxies). This *context-sensitivity* of computer simulations' phenomenology at the meso-scale is a powerful reminder of the epistemic limits of computer simulations over ordinary experiments (much more than any debate about the materiality of experiments over the computer simulations). The problem with computer simulations in contemporary cosmology has nothing to do with the non-materiality of the target system, as philosophers of science have been debating. Instead, it has to do with the sensitivity of the simulations to a plurality of contextual causal factors, whose stochastic nature and specific counteracting or contributing role in individual galaxies' formation is quite simply impossible to factor in precisely. This context-sensitivity of computer simulations ultimately impacts the ability of ΛCDM to provide satisfactory *causal explanations* for the precise astrophysical data and measurements behind BTF and MDAR. If, despite these epistemic limits, the simulations are still capable of retrieving the BTF relation, this is success enough, and must count as success enough for ΛCDM—most cosmologists would conclude. Agreed. It is also agreed that the future of the debate between ΛCDM and MOND does not particularly hang on these computer simulations. Nevertheless, the success of computer simulations at the meso-scale is also a powerful reminder that curve-fitting is not tantamount to giving an *explanation* for the BTF and MDAR relations. These remain open (and in my view non-negligible) epistemological questions for ΛCDM to answer.

## 4. The upscaling problem

The upscaling problem is the opposite of the downscaling problem and it affects Milgrom's MOND as a prominent rival to the ΛCDM model at the meso-scale. *The upscaling problem* consists in MOND's ability to *consistently* retrieve large-scale structures, such as galaxy clusters, and structure formation more in general at a scale where general relativity applies. MOND has proved to work really well to fit galaxy rotation curves and is regarded as a very successful phenomenology at the meso-scale. The BTF relation follows from MOND and MDAR is also naturally explained by MOND. However, despite the success at the meso-scale, MOND is by and large regarded as just that: a successful phenomenology for galaxy's rotation curves. The theory does not fare equally well when it comes to large-scale phenomena such as explaining galaxy formation from primordial fluctuations (anisotropies) in the cosmic microwave background, or explaining galaxy clusters and large scale structure formation more in general—if there is no dark matter, and with no general relativity either, how to explain these pieces of evidence at the large scale?

Galaxy clusters have a very high internal gravitational acceleration and the effects of MOND are too weak to explain it. Within MOND this recalcitrant evidence at the level of galaxy clusters has been met with mixed feelings. Some have argued that there might be non-luminous baryonic matter or neutrinos (see Angus, Famaey, & Buote, 2008; Milgrom, 2008) to explain the problem at the galaxy cluster level; others have seen the negative evidence at large-scale as an argument that MOND is incomplete (but not necessarily incorrect). In what follows, for limits of space, I am going to concentrate on the latter option and give a very brief overview of some recent work within the MOND paradigm to address the upscaling problem (i.e. how to retrieve galaxy clusters phenomenology within MOND).

One recent proposal goes under the name of Extended MOND or EMOND (Zhao & Famaey, 2012). The idea is to make the MOND acceleration scale $a_0$ a function of the gravitational potential $A_0(\Phi)$ so as to be able to retrieve higher accelerations for galaxy clusters within EMOND. In other words, the acceleration scale $a_0$ would not be constant (as in MOND) but instead depending on the distance from the deep potential well, the galaxy cluster would have a different acceleration: much higher than $a_0$ in regions of the galaxy cluster closer to the deep potential well. The clear advantage of introducing this new feature is that EMOND has the potential to mimic dark matter-like effects at high accelerations in galaxy clusters. The idea has been tested in the so-called Ultra-Diffuse Galaxies (UDG) in the Coma Cluster (Hodson & Zhao, 2017). UDG contain very little gas and seem to be composed primarily of dark matter. But the applicability is still very limited in part because UDG are little studied (with very small sample size as of today) to afford robust statistical results. More to the point, significant assumptions are also made within EMOND. One of them, expressly acknowledged by EMOND proponents, is that to get an estimate for the dynamical and stellar mass of the UDG, EMOND resorts to a technique called fundamental manifold (FM) from which it is possible to calculate the velocity dispersions, and from there it is possible to calculate the dynamical mass of the UDG. Yet, FM under-predicts the velocity dispersion and to mitigate this discrepancy, adjustments have to be made in EMOND boundary potential for the Coma cluster. More to the point, EMOND does not give yet any explanation or mechanism for galaxy cluster formation. For example, evidence coming from the Bullet Cluster[18] suggests that the missing mass in clusters (i.e. what Zwicky originally called dark matter) must be collisionless. And the collisionless nature of the missing mass in clusters is and remains outside the explanatory scope of EMOND (and MOND more in general).[19]

---

[18] The Bullet Cluster (i.e. galaxy cluster 1E 0657-56) is the merger of two galaxy clusters that provides key evidence for the existence of dark matter (see Clowe, Gonzalez, & Markevitch, 2004). The intergalactic gas that collided emitted X-rays that can be observed. But using lensing it is possible to see that most of the mass of the galaxy merger is not in fact located in the intra-cluster matter, but around it (displaying a distinctive halo around the merged clusters). The Bullet Cluster has been used to fix constraints on possible particle models for dark matter because it suggests that dark matter is somehow collisionless. One such particle candidate for collisionless cold dark matter is the so-called self-interacting dark matter (SIDM). For a recent review of how N-body simulations of collisionless cold dark matter fare vis-à-vis the observation from the Bullet Cluster, see Robertson, Massey, and Eke (2017).

[19] Famaey and McGaugh (2012), Section 6.6.4. review five possible options within MOND to deal with the problem of missing mass in clusters: "(i) Practical falsification of MOND; (ii) Evidence for missing baryons in the central parts of clusters; (iii) Evidence for non-baryonic dark matter (existing or exotic); (iv) Evidence that MOND is an incomplete paradigm; (v) Evidence for the effect of additional fields in the parent relativistic theories of MOND, not included in Milgrom's formula" p. 82. For reasons of space I cannot discuss these different options and I refer the reader to Famaey and McGaugh's article. Suffice to say that the existence of some unseen form of mass (exotic neutrinos, for example) is also contemplated in some MOND versions for clusters.

Thus, as with the downscaling problem, here too we have an example of a theory (EMOND), which has the potential of retrieving some dark-matter-like effects in particular kinds (UDG) of galaxy clusters. For now, the data on UGD are too limited and the theoretical progress made is not quite advanced enough to successfully upgrade EMOND from its status of being a successful phenomenology at the meso-scale.

Moreover, large-scale structure is and remains the biggest 'upscaling problem' for MOND. For MOND is nowhere nearer to providing an explanation or understanding of the formation and evolution of structure at large scale, where ΛCDM has traditionally held a stronghold. Thus, any future variations of MOND will have to address the problem of delivering at the large-scale structure and at the level of galaxy clusters. Clearly, MOND cannot be blamed for being what it is not: MOND as modified Newtonian dynamics does not assume general relativity, upon which the explanation of large-scale structure formation relies, as discussed in Section 2. Yet, the burden is on MOND to come up with an alternative (MOND-like) way of calculating for example the CMB angular power spectrum. This is ultimately the "upscaling problem" for MOND: how to *consistently* develop a relativistic version of MOND that is much needed to recover phenomena at the large scale.

Some tentative and preliminary suggestions to this effect have been made. For example, it has been shown that the amplitude ratio of the first to the second peak in the power spectrum of temperature fluctuations in the CMB could be in principle predicted a priori from a very simple *ansatz* general relativity without cold dark matter (McGaugh, 1999 and 2000). This was done well before precise data from Planck (Ade et al., 2014) became available to constrain such predictions about CMB fluctuations (the first data were from Bernardis, Ade et al., 2000). But even such an *ansatz* (which in itself did not amount to anything like a possible relativized MOND) was not able to retrieve the third peak and any following ones (for a review of this debate see McGaugh, 2015b). By contrast, ΛCDM fits the data about the CMB power spectrum very well (for WMAP data fit see Komatzu, Smith et al., 2011). Thus, in the absence of an alternative (MOND-like) way of predicting the CMB power spectrum, ΛCDM remains the only paradigm to explain large-scale structure.

Before we conclude this section, another recent proposal within the family of MOND is worth mentioning (although very briefly, for reasons of space). Verlinde's (2016) recent proposal on emergent gravity has attracted significant attention in the community.[20] Taking as its starting point quantum information and entanglement entropy, Verlinde has argued that thermal excitations responsible for the de Sitter entropy may be regarded as constituting positive dark energy. In other words, dark energy and the accelerated expansion of the universe could be regarded as the product of what Verlinde refers to as the slow thermalization of the emergent spacetime. The idea is that at small scale (smaller than the Hubble radius gravity) general relativity applies. But at large scale the de Sitter entropy and slow thermalization leads to deviations from GR and to emergent gravity. In Verlinde's own words (Verlinde's (2016), p. 6), "the volume law contribution to the entanglement entropy, associated with the positive dark energy, turns the otherwise 'stiff' geometry of spacetime into an elastic medium. We find that the elastic response of this 'dark energy' medium takes the form of an extra 'dark' gravitational force that appears to be due to 'dark matter'." An intriguing idea underlies Verlinde's proposal; namely, that it is logically possible that the laws which govern the long time and distance scale dynamics in our universe are decoupled from the emergent local laws of physics described by our current effective field theories …. our universe contains a large amount of quantum information in extremely long range correlations of the underlying microscopic degrees of freedom. The present local laws of physics are not capable of detecting or describing these localized states. (Verlinde's (2016), p. 13).

Obviously to be logically possible is not one and the same as to be physically possible, and the challenge ahead for emergent gravity is to provide a clear physical mechanism for this envisaged decoupling of the laws that govern long distance scale dynamics in the universe from the local laws described by current effective field theories. Verlinde introduces the analogy with a glassy system where at short observation times a glassy system is indistinguishable from a crystalline system and their effective field theory description would be the same. But at long timescale their respective dynamics would differ remarkably, and the long-distance scale dynamics of a glassy system cannot be derived from the effective description of the short distance behavior.

Dark matter features in Verlinde's theory as 'apparent': in a de Sitter spacetime matter would create a memory effect in the dark energy medium by removing entropy from an inclusion region. This would produce an elastic stress in the medium, and in turn cause a reaction force on the matter. This quasi-elastic force is behind the "excess gravity that is currently attributed to dark matter" (Verlinde's (2016) p. 14). The Tully-Fisher relation can be derived from it, which is an important result of Verlinde's proposal (Verlinde's (2016), pp. 23ff). Verlinde is adamant in stressing the difference between his own proposal and pure MOND as follows:

In our description there is no modification of the law of inertia, nor is our result (7.43—[BTF]) to be interpreted as a modified gravitational field equation. It is derived from an estimate of an effect induced by the displacement of the free energy of the underlying microscopic state of the de Sitter space due to matter. (…). Hence, although we derived the same relation as modified Newtonian dynamics, the physics is very different. … There is little dispute about the observed scaling relation (7.43), but the disagreement in the scientific community has mainly been about whether it represents a new law of physics. In our description it does not. (Verlinde's (2016), p. 39).

When it comes to possible experimental tests for emergent gravity, some provisional positive results from weak gravitational lensing were published in December 2016 (see Brouwer et al., 2017). The test measured the average surface mass density profiles of a large sample of isolated central galaxies taken from the photometric Kilo-Degree Survey (KiDS) and the spectroscopic GAMA survey (isolated GAMA galaxies without satellites). This sample satisfied as much as possible Verlinde's idealizations that galaxies should be sufficiently isolated and spherically symmetric. Brouwer et al. have found a good agreement between the available data for the chosen sample and the predictions made by Verlinde's emergent gravity: the emergent-gravity predicted profiles follow closely that of virialized systems that typically obey NFW profile for dark matter.

But the provisional positive news coming from the weak lensing test has been counterbalanced by a stream of more cautious experimental results that have followed in the first months of 2017. Here are some highlights. Ettori et all. (2017) have tested emergent gravity with data coming from massive X-ray luminous galaxy clusters Abell 2142 and Abell 2319, which also seem to follow some of Verlinde's idealizations because they are almost spherical, and quite isolated (with no major neighbor object). And they found that emergent gravity does not fare well in the inner regions of the galaxy cluster ($r < 200$ kpc). Equally more cautious conclusions have come out of Hees et al. (2017) study concerning the ability of emergent gravity to reproduce MOND in galaxies and in the solar system. At galaxy level, for galaxy rotation curves, Hees et al. found that emergent gravity produces less good quality fits than MOND, with the best fits being at low regimes.

---

[20] Verlinde's emergent gravity, by contrast with other MOND proposals, has a theoretical basis although it shares similarities with Milgrom's (1999) modified inertia proposal. Hence the rationale for including it in this Section. I thank a reviewer for pressing me on this point.

Lelli, McGaugh, and Schombert (2017) have also published results showing that the equivalence between MOND and emergent gravity applies only for point mass approximations, but fails when dealing with more realistic scenarios of finite-size galaxies. The predicted galaxies' rotation curves in Emergent Gravity display a systematic hook-shape deviation when compared to ca. 2700 data points from 153 disc galaxies. But perhaps the main challenge to date on Emergent Gravity comes from the same proponents of MOND. Milgrom and Sanders (2016) have launched six main objections against Verlinde's view:

- Shaky theoretical grounds: "the idea is not yet based on some underlying, full-fledged, microscopic theory" (Milgrom and Sanders (2016), p. 2).
- Unjustified idealizations: the *Ansatz* used by EG (e.g. that the entanglement entropy of de Sitter space is distributed over the entire universe) is plucked out of thin air.
- Too many abstractions: emergent gravity ignores possible contributions to the gravitational potential $\Phi(r)$ coming from possible baryonic mass outside the radius $r$.
- Non-universality: it is not clear how to move beyond the assumption of spherically symmetric, isolated system to more realistic scenarios concerning finite-size galaxies.
- Theoretical parasitism: emergent gravity seems to be designed to reproduce MOND phenomenology "and so ride on its successes" (Milgrom and Sanders (2016), p. 2).
- Empirical equivalence with MOND: Brouwer et al. (2017) weak lensing analysis tests effectively the many-galaxies-average gravitation potential at large radii, and since the radii are so large, they can be approximated by a point mass. So, if anything, Brouwer et al. is a confirmation of MOND predictions insofar as Verlinde rides on MOND's success.

Obviously, this is a cutting-edge debate that is very much open-ended and ongoing. Thus, all that one can say and conclude here is that more work needs be done to fully appreciate the theoretical novelty of Emergent Gravity, the extent to which it might or might not really differ from MOND and its empirical viability.

## 5. The 'in between' scales problem

To address the problem of cosmological modelling at the meso-scale, very recently some physicists have endeavored to reconcile ΛCDM with MOND by exploring alternative solutions, which are meant to achieve the best of both worlds at the meso scale. This is a fascinating area of inquiry where some recent creative attempts at modelling 'in between' scales have been proposed, which try to retain the successful phenomenology of MOND at the meso-scale, while also introducing dark matter in the framework to retrieve large-scale structure formation. For reasons of space, I confine my attention to a very brief discussion of one prominent recent example of this kind of 'in between' scales modelling by Justin Khoury and collaborators at Penn.[21]

The work of Berezhiani and Khoury (Berezhiani & Khoury, 2015; Khoury, 2016) is sometimes referred to as GMOND, or generalized MOND (see Hodson & Zhao, 2017) and revolves around the idea of 'dark matter superfluidity'. The central idea is to reconcile dark matter and MOND by treating them as a Janus-faced entity: they would represent different phases of a single underlying superfluid substance, whereby dark matter would behave like a superfluid (with no entropy and vanishing viscosity) inside galaxies where MOND's successful phenomenology applies. But superfluidity would break down (reverting to a normal fluid carrying entropy and viscosity at high temperature) at the large scale of galaxy clusters where ΛCDM traditionally fares much better than MOND.

This Janus-faced nature is explained in terms of dark matter halos having a velocity that translates into temperature. In individual galaxies, where the halo velocity is slow, the temperature of dark matter would be relatively low (ca. 0.1 mK), which would explain the superfluid behavior that is compatible with MOND treatment of galaxy rotational curves. But in galaxy clusters, where the velocity is higher, and hence the temperature is higher (ca. 10 mK), superfluidity would break down and default dark matter mechanisms (as per ΛCDM) would ensue. This proposal would seem to have the potential of solving the problem of modelling at the meso-scale in cosmology without having to face either the downscaling problem typical of HYDs and SAMs in ΛCDM; nor the upscaling problem currently faced by MOND and EMOND. The key idea is to start with the empirical facts about BTF and MDAR— as opposed to empirical facts about large scale structure—and work out dark matter scenarios that might fit those empirical facts in the first instance. The originality of Khoury and collaborators' hybrid approach consists precisely in the central role played by empirical evidence at the meso-scale (BTF and MDAR) in rethinking the ΛCDM–MOND debate.

However, the price to pay for this 'in between' scales modelling is that in the superfluid phase, dark matter cannot be described by particles such as WIMPS that are collisionless, and as such do not thermalize. How to reconcile dark matter superfluidity, which thermalizes at high temperature, with current searches for collisionless dark matter candidates in LUX, PandaX-II and other direct detection experiments—remains to be seen. There might be a genuine risk of successfully bridging the gap between meso and large-scale, at the cost of creating a new tension between the meso-scale and the micro-scale of putative dark matter particle candidates.

With an eye to addressing this issue, Berezhiani, Famaey, and Khoury (2017) have proposed particle physics models for dark matter superfluidity that do not incur into the aforementioned problem about thermalization. For example, instead of collisionless WIMPs, they assume that the DM superfluid might consist of axion-like particles with strong self-interactions that Bose-Einstein condense in the galaxy halos. Dark matter would be better described in terms of collective excitations (phonons) that would couple with baryons. Using this hypothesis (albeit slightly revised compared to Berezhiani & Khoury, 2015 so that superfluidity would now make up only a small fraction of the DM halo), Berezhiani, Famaey, and Khoury (2017) have tried to match NFW dark matter halos and retrieve galaxies's rotations curves with mixed results.[22]

---

[21] Another example of an 'in between' approach is Blanchet and Le Tiec (2008), who hypothesized a model of dark matter and dark energy as a dipolar medium, with a dipole moment vector polarizable in a gravitational field and showed how this dipolar fluid could reproduce ΛCDM at cosmological scale as well as MOND at galactic scale.

[22] Two representative galaxies were chosen as test cases, with the IC2574 being a recalcitrant one for ΛCDM supplemented by feedback. But testable phenomenological consequences for more complex systems, such as galaxy clusters, dwarf satellites and ultra-diffuse galaxies, are still at a qualitative level.

More recently, Famaey, Khoury, and Penco (2017) have further addressed the problem of bridging the gap between DM superfluidity at the meso scale with DM particle models at the micro scale. Taking once again MDAR as a hard empirical fact at the meso scale, which allows to uniquely predict the DM density profile of galaxies from their baryonic distribution, the new model explores how such empirically known distribution of dark matter could possibly arise from particle physics interactions between dark matter and ordinary matter.[23] The desired DM distribution would arise as an equilibrium solution thanks to DM strong interactions with ordinary matter (where again DM cannot be WIMPs in this model and is assumed to exchange energy with baryons via elastic collisions). The bonus of this new model is that it shows how this particle physics model could naturally lead disk galaxies to equilibrium configurations that match the MDAR , without any need to invoke ad hoc feedback mechanisms. Galaxy clusters, traditionally a stronghold of ΛCDM, would not follow MDAR because they would not reach equilibrium configurations in this model.

These are exciting and promising ideas. Yet, some worries remain. First, it is still unclear how superfluid dark matter can provide an explanation for the formation of large-scale structure from primordial fluctuations in the CMB; or how it can retrieve all the empirical successes of ΛCDM, in explaining BBN and BAO matter spectrum, for example. The deliberate choice of placing BTF and MDAR center stage and work out particle physics models at the micro scale that could give rise to the right meso-scale structure, faces itself a version of the 'upscaling' problem. Another worry one might have is about the predictive scope of DM superfluidity. To what extent do the more recent positive results (in retrieving for example galaxy IC2574's rotation curve) piggy-back on the fact that superfluid core is assumed to make up only a small fraction of the DM halo? Is not DM superfluidity successful to the extent that dark matter is successful at predicting the relevant phenomena? Thus, the 'in between' scales problem affecting hybrid models is ultimately a problem about *predictive novelty*: to what extent the hybrid proposal is able to predict novel phenomena rather than piggy-back on the existing predictive and explanatory power of both ΛCDM at large scale and MOND at the meso scale. More work on superfluidity numerical simulations is currently under way and the jury remains out.

## 6. Concluding remarks

To conclude, what is to be said about multi-scale modelling in contemporary cosmology? Going back to Batterman's (2013) comment, with which I opened this paper, i.e. that "mesoscopic structures cannot be ignored and, in fact, provide the bridges that allow us to model across scales", as recent developments in cosmology clearly show, the meso scale (i.e. the scale of individual galaxies) prove indeed crucial in successfully modelling across scales. The current verdict on ΛCDM over MOND depends crucially on their respective ability of modelling across scales, and dealing with some of the specific problems that arise along the way. The main upshot of this article has been to offer an introduction to very recent ongoing research in this fascinating area and present three main problems facing multi-scale modeling in contemporary cosmology.

The ΛCDM model, which is the received view and the most successful of the current cosmological models, faces nonetheless the downscaling problem when it comes to deliver on MDAR and BTF. This is a problem about the ability of ΛCDM to *causally explain* (and not just retrieve via simulations and ad hoc feedback mechanisms) these two phenomena. While the fast-growing development of HYD and SAM simulations has addressed this problem and has been able to retrieve MDAR and BTF, worries still linger about some of the epistemological assumptions behind these computer simulations. As discussed in Section 3, computer simulations within ΛCDM resort to feedback, and how feedback enters into and is modelled in different kinds of simulations (so as to successfully retrieve the BTF) raises interesting methodological questions about the nature and limits of the idealizations used in HYDs and SAMs.

The upscaling problem affects MOND and its ability to *consistently* retrieve large-scale observations (like CMB angular power spectrum, structure formation, galaxy clusters and even lensing) at a scale where general relativity applies. Recent attempts at extending MOND (EMOND) have had a limited empirical success so far, and are still far from being able to even provide a physical explanation for possible formation mechanisms for galaxy clusters and large-scale structure. Verlinde's emergent gravity, and its success at deriving BTF has been regarded by MOND supporters as simply reliant on MOND rather than a genuine theory in its own right.

Finally, the 'in between' scales problem affects proposals designed to achieve the best of both worlds at the meso-scale. This is a fascinating area from a physical and a philosophical point of view, where the main challenge is the ability to offer genuine *predictive novelty* over and above the mixing-and-matching of successful features of ΛCDM and MOND.

Modelling at the meso scale has to be *explanatory powerful, theoretically consistent (across large scales)* and *predictively novel* to offer genuine bridges between the micro-scale and the large-scale. Answers to pressing questions about the very existence and nature of dark matter depend not just on detecting dark matter particles in the laboratory, but also on our ability to find answers to these problems about multi-scale modelling.

## Acknowledgements

Earlier versions of this paper were presented at the Rotman Institute of Philosophy conference on Philosophy of Cosmology, the Physics of Fine-Tuning conference in Crete and at the Royal Observatory of Edinburgh. I thank Chris Smeenk, Sarah Gallagher, Wayne Myrvold, Tessa Baker, Lee Smolin, Daniele Oriti, Michael Hicks, Roger Davies, Ken Rice, Agnès Ferté and Joe Zunzt and the audiences for helpful questions and comments. I am very grateful to Alex Murphy, John Peacock, Stacy McGaugh, Justin Khoury, and Andrea Cattaneo for very constructive comments on earlier versions of this paper. This article is part of a project that has received funding from the European Research Council under the European Union's Horizon 2020 research and innovation programme (grant agreement European Consolidator Grant H2020-ERC-2014-CoG 647272 *Perspectival Realism. Science, Knowledge, and Truth from a Human Vantage Point*).

---

[23] I thank Khoury for helpful correspondence on this latest research.